\documentstyle[12pt,twoside]{article}

\def\a{\alpha}
\def\b{\beta}
\def\c{\chi}
\def\d{\delta}
\def\e{\epsilon}                
\def\f{\phi}                    
\def\g{\gamma}
\def\h{\eta}

\def\j{\psi}

\def\l{\lambda}
\def\m{\mu}
\def\n{\nu}

\def\p{\pi}                     
\def\r{\rho}                    
\def\s{\sigma}                  
\def\t{\tau}

\def\x{\xi}

\def\D{\Delta}
\def\F{\Phi}

\def\L{\Lambda}

\def\S{\Sigma}

\def\ch{{\cal H}}

\def\cl{{\cal L}}

\catcode`@=11

\def\un#1{\relax\ifmmode\@@underline#1\else $\@@underline{\hbox{#1}}$\relax\fi}

{\catcode`\'=\active \gdef'{{}~\bgroup\prim@s}}

\def\magstep#1{\ifcase#1 \@m\or 1200\or 1440\or 1728\or 2074\or 2488\or
        2986\fi\relax}

\font\twfvmi=cmmi10\@magscale5
    \skewchar\twfvmi='177
\font\twfvsy=cmsy10\@magscale5
    \skewchar\twfvsy='60
\font\twfvly=lasy10\@magscale5
\font\thtyrm=cmr10\@magscale6

\def\vpt{\textfont\z@\fivrm
  \scriptfont\z@\fivrm \scriptscriptfont\z@\fivrm
\textfont\@ne\fivmi \scriptfont\@ne\fivmi \scriptscriptfont\@ne\fivmi
\textfont\tw@\fivsy \scriptfont\tw@\fivsy \scriptscriptfont\tw@\fivsy
\textfont\thr@@\tenex \scriptfont\thr@@\tenex \scriptscriptfont\thr@@\tenex
\def\prm{\fam\z@\fivrm}%
\def\unboldmath{\everymath{}\everydisplay{}\@nomath
  \unboldmath\fam\@ne\@boldfalse}\@boldfalse
\def\boldmath{\@subfont\boldmath\unboldmath}%
\def\pit{\@getfont\pit\itfam\@vpt{cmti5}}%
\def\psl{\@subfont\sl\it}%
\def\pbf{\@getfont\pbf\bffam\@vpt{cmbx5}}%
\def\ptt{\@subfont\tt\rm}%
\def\psf{\@subfont\sf\rm}%
\def\psc{\@subfont\sc\rm}%
\def\ly{\fam\lyfam\fivly}\textfont\lyfam\fivly
    \scriptfont\lyfam\fivly \scriptscriptfont\lyfam\fivly
\@setstrut\rm}

\def\@vpt{}

\def\vipt{\textfont\z@\sixrm
  \scriptfont\z@\sixrm \scriptscriptfont\z@\sixrm
\textfont\@ne\sixmi \scriptfont\@ne\sixmi \scriptscriptfont\@ne\sixmi
\textfont\tw@\sixsy \scriptfont\tw@\sixsy \scriptscriptfont\tw@\sixsy
\textfont\thr@@\tenex \scriptfont\thr@@\tenex \scriptscriptfont\thr@@\tenex
\def\prm{\fam\z@\sixrm}%
\def\unboldmath{\everymath{}\everydisplay{}\@nomath
  \unboldmath\@boldfalse}\@boldfalse
\def\boldmath{\@subfont\boldmath\unboldmath}%
\def\pit{\@subfont\it\rm}%
\def\psl{\@subfont\sl\rm}%
\def\pbf{\@getfont\pbf\bffam\@vipt{cmbx6}}%
\def\ptt{\@subfont\tt\rm}%
\def\psf{\@subfont\sf\rm}%
\def\psc{\@subfont\sc\rm}%
\def\ly{\fam\lyfam\sixly}\textfont\lyfam\sixly
    \scriptfont\lyfam\sixly \scriptscriptfont\lyfam\sixly
\@setstrut\rm}

\def\@vipt{}

\def\xxxpt{\textfont\z@\thtyrm
  \scriptfont\z@\twfvrm \scriptscriptfont\z@\twtyrm
\textfont\@ne\twfvmi \scriptfont\@ne\twfvmi \scriptscriptfont\@ne\twtymi
\textfont\tw@\twfvsy \scriptfont\tw@\twfvsy \scriptscriptfont\tw@\twtysy
\textfont\thr@@\tenex \scriptfont\thr@@\tenex \scriptscriptfont\thr@@\tenex
\def\unboldmath{\everymath{}\everydisplay{}\@nomath\unboldmath
        \textfont\@ne\twfvmi \textfont\tw@\twfvsy \textfont\lyfam\twfvly
        \@boldfalse}\@boldfalse
\def\boldmath{\@subfont\boldmath\unboldmath}%
\def\prm{\fam\z@\thtyrm}%
\def\pit{\@subfont\it\rm}%
\def\psl{\@subfont\sl\rm}%
\def\pbf{\@getfont\pbf\bffam\@xxxpt{cmbx10\@magscale6}}%
\def\ptt{\@subfont\tt\rm}%
\def\psf{\@subfont\sf\rm}%
\def\psc{\@subfont\sc\rm}%
\def\ly{\fam\lyfam\twfvly}\textfont\lyfam\twfvly
   \scriptfont\lyfam\twfvly \scriptscriptfont\lyfam\twtyly
\@setstrut \rm}

\def\@xxxpt{}

\def\Huge{\@setsize\Huge{36pt}\xxxpt\@xxxpt}

\font\thtymi=cmmi10\@magscale6
    \skewchar\thtymi='177
\font\thtysy=cmsy10\@magscale6
    \skewchar\thtysy='60
\font\thtyly=lasy10\@magscale6
\font\thsirm=cmr12\@magscale6

\def\xxxvipt{\textfont\z@\thsirm
  \scriptfont\z@\thtyrm \scriptscriptfont\z@\twfvrm
\textfont\@ne\thtymi \scriptfont\@ne\thtymi \scriptscriptfont\@ne\twfvmi
\textfont\tw@\thtysy \scriptfont\tw@\thtysy \scriptscriptfont\tw@\twfvsy
\textfont\thr@@\tenex \scriptfont\thr@@\tenex \scriptscriptfont\thr@@\tenex
\def\unboldmath{\everymath{}\everydisplay{}\@nomath\unboldmath
        \textfont\@ne\thtymi \textfont\tw@\thtysy \textfont\lyfam\thtyly
        \@boldfalse}\@boldfalse
\def\boldmath{\@subfont\boldmath\unboldmath}%
\def\prm{\fam\z@\thsirm}%
\def\pit{\@subfont\it\rm}%
\def\psl{\@subfont\sl\rm}%
\def\pbf{\@getfont\pbf\bffam\@xxxpt{cmss12\@magscale6}}%
\def\ptt{\@subfont\tt\rm}%
\def\psf{\@subfont\sf\rm}%
\def\psc{\@subfont\sc\rm}%
\def\ly{\fam\lyfam\thtyly}\textfont\lyfam\thtyly
   \scriptfont\lyfam\thtyly \scriptscriptfont\lyfam\twfvly
\@setstrut \rm}

\def\@xxxvipt{}

\def\HUGE{\@setsize\HUGE{43pt}\xxxvipt\@xxxvipt}

\catcode`@=12

\font\tenex=cmex10 scaled 1200

\def\Sc#1{\hbox{\sc #1}}        
\font\oo=lcirclew10            

\def\bo{{\raise.05ex\hbox{\large$\Box$}\:}}             
\def\cbo{{\,\raise-.15ex\Sc [\,}}                       
\def\pa{\partial}                                       
\def\su{\sum}                                           
\def\TH{{\raise.2ex\hbox{$\displaystyle \bigodot$}\mskip-4.7mu \llap H \;}}
\def\face{\hbox{\normalsize$\;\;\:{\raise.9ex\hbox{\oo n}\mskip-13mu \llap
        {${\buildrel{\hbox{\frtnrm ..}}\over\smile}$}}\:$}}     
\def\Face{{\raise.2ex\hbox{$\displaystyle \bigodot$}\mskip-2.2mu \llap {$\ddot
        \smile$}}}                                      
\def\Lhat{{\bf\rlap{\kern-.09em$\hat{\phantom L}$}L}}
\def\Lcheck{{\bf\rlap{\kern-.09em$\check{\phantom L}$}L}}

\def\sp#1{{}^{#1}}                              
\def\sb#1{{}_{#1}}                              
\def\sl#1{\rlap{\hbox{$\mskip 1 mu /$}}#1}      
\def\sket#1{\left| #1\right\rangle}             
\def\sVEV#1{\left\langle #1\right\rangle}       
\def\leftrightarrowfill{$\mathsurround=0pt \mathord\leftarrow \mkern-6mu
        \cleaders\hbox{$\mkern-2mu \mathord- \mkern-2mu$}\hfill
        \mkern-6mu \mathord\rightarrow$}
\def\dvec#1{\vbox{\ialign{##\crcr
        \leftrightarrowfill\crcr\noalign{\kern-1pt\nointerlineskip}
        $\hfil\displaystyle{#1}\hfil$\crcr}}}           
\def\dt#1{{\buildrel {\hbox{\LARGE .}} \over {#1}}}     
\def\ddt#1{{\buildrel {\hbox{\LARGE .\kern-2pt.}} \over {#1}}}

\def\frac#1#2{{\textstyle{#1\over\vphantom2\smash{\raise.20ex
        \hbox{$\scriptstyle{#2}$}}}}}                   
\def\ha{\frac12}                                        
\def\sfrac#1#2{{\vphantom1\smash{\lower.5ex\hbox{\small$#1$}}\over
        \vphantom1\smash{\raise.4ex\hbox{\small$#2$}}}} 
\def\bfrac#1#2{{\vphantom1\smash{\lower.5ex\hbox{$#1$}}\over
        \vphantom1\smash{\raise.3ex\hbox{$#2$}}}}       
\def\afrac#1#2{{\vphantom1\smash{\lower.5ex\hbox{$#1$}}\over#2}}    

\def\boxes#1{
        \newcount\num
        \num=1
        \newdimen\downsy
        \downsy=-1.64ex
        \mskip-7.8mu
        \bo
        \loop
        \ifnum\num<#1
        \llap{\raise\num\downsy\hbox{$\bo$}}
        \advance\num by1
        \repeat}
\def\boxup#1#2{\newcount\numup
        \numup=#1
        \advance\numup by-1
        \newdimen\upsy
        \upsy=.82ex
        \mskip7.8mu
        \raise\numup\upsy\hbox{$#2$}}

\newskip\humongous \humongous=0pt plus 1000pt minus 1000pt
\def\caja{\mathsurround=0pt}

\newif\ifdtup
\def\panorama{\global\dtuptrue \openup2\jot \caja
        \everycr{\noalign{\ifdtup \global\dtupfalse
        \vskip-\lineskiplimit \vskip\normallineskiplimit
        \else \penalty\interdisplaylinepenalty \fi}}}
\def\li#1{\panorama \tabskip=\humongous                         
        \halign to\displaywidth{\hfil$\displaystyle{##}$
        \tabskip=0pt&$\displaystyle{{}##}$\hfil
        \tabskip=\humongous&\llap{$##$}\tabskip=0pt
        \crcr#1\crcr}}

\def\NP{Nucl. Phys. B}
\def\PL{Phys. Lett. }

\def\PRD{Phys. Rev. D}
\def\ref#1{$\sp{#1]}$}

\parskip=\medskipamount                 
\def\baselinestretch{1.2}       
\thicklines                         
\def\title#1#2#3#4{
\begin{document}
        {\hbox to\hsize{#4 \hfill QMW/PH/ #3}}\par
        \begin{center}\vskip.5in minus.1in {\Large\bf #1}\\[.5in minus.2in]{#2}
        \vskip1.4in minus1.2in {\bf ABSTRACT}\\[.1in]\end{center}
        \begin{quotation}\par}
\def\author#1#2{#1\\[.1in]{\it #2}\\[.1in]}
\def\AM{Aleksandar Mikovi\'c\,\footnote
   {Work supported by the U.K. Science and Engineering Research Council}
\footnote{E-mail address: MIKOVIC@V1.PH.QMW.AC.UK}
\\[.1in] {\it Department of Physics, Queen Mary and Westfield
College,\\ Mile End Road, London E1 4NS, U.K.}\\[.1in]}
\def\endtitle{\par\end{quotation}\vskip3.5in minus2.3in\newpage}
\def\endabstract{\par\end{quotation}
        \renewcommand{\baselinestretch}{1.2}\small\normalsize}

\def\sect#1{\bigskip\medskip\goodbreak\noindent{\large\bf{#1}}\par\nobreak
        \medskip\markright{#1}}
\def\chsc#1#2{\phantom m\vskip.5in\noindent{\LARGE\bf{#1}}\par\vskip.75in
        \noindent{\large\bf{#2}}\par\medskip\markboth{#1}{#2}}
\def\Chsc#1#2#3#4{\phantom m\vskip.5in\noindent\halign{\LARGE\bf##&
        \LARGE\bf##\hfil\cr{#1}&{#2}\cr\noalign{\vskip8pt}&{#3}\cr}\par\vskip
        .75in\noindent{\large\bf{#4}}\par\medskip\markboth{{#1}{#2}{#3}}{#4}}
\def\chap#1{\phantom m\vskip.5in\noindent{\LARGE\bf{#1}}\par\vskip.75in
        \markboth{#1}{#1}}
\def\refs{\bigskip\medskip\goodbreak\noindent{\large\bf{REFERENCES}}\par
        \nobreak\bigskip\markboth{REFERENCES}{REFERENCES}
        \frenchspacing \parskip=0pt \renewcommand{\baselinestretch}{1}\small}
\def\unrefs{\normalsize \nonfrenchspacing \parskip=medskipamount}
\def\Item{\par\hang\textindent}
\def\Itemitem{\par\indent \hangindent2\parindent \textindent}
\def\makelabel#1{\hfil #1}
\def\topic{\par\noindent \hangafter1 \hangindent20pt}
\def\Topic{\par\noindent \hangafter1 \hangindent60pt}

\title{Canonical Quantization of 2d Gravity Coupled to $c<1$ Matter}
{\AM}{91/22}{February 1992}
We study 2d gravity coupled to $c<1$ matter through canonical quantization
of a free scalar field, with background charge, coupled to
gravity. Various features of the theory can be more easily understood
in the canonical approach, like gauge indipendence of the path-integral
results and the absence of the local physical degrees of freedom.
By performing a non-canonical transformation of the phase space variables,
we show that the theory takes a free-field form, i.e. the constraints become
the free-field Virasoro constraints. This implies that the David-Distler-Kawai
results can be derived in a gauge indipendent way,
and also proves the free-field assumption which was used
for obtaining the spectrum of the theory in the conformal gauge.
A discussion of the physical spectrum of the theory is presented, with an
analysis of the unitarity of the discrete momentum states.

\endtitle

\sect{1. Introduction}

Understanding the two dimensional quantum gravity is important
because of its relevance for the non-critical string theory, statistical
mechanics of random surfaces, and as a toy model of
quantum gravity in four dimensions. The
theory so far has been mainly analyzed in the path-integral quantization
scheme \cite{{kpz},{ddk}}.
Although many important results have been achived in this scheme, it is
also important to understand the theory in the Dirac canonical quantization
approach \cite{dir}.
First, the path integral quantization of a gauge invariant system
requires gauge fixing, so that the questions of
gauge indipendence and relation of the
results in different gauges inevitably appear. This has been automaticaly
taken care of in the Dirac approach, since it is a gauge indipendent
quantization method. Second, understanding the relation between the
path-integral and the Dirac quantization results
is important question in its own right,
especially if one is interested in possible relations to
4d quantum gravity, where the
Dirac quantization iz much better understood than the path-integral
quantization.

The first exact results in 2d quantum gravity
were obtained by Knizhnik, Polyakov and Zamolodchikov (KPZ),
who studied 2d quantum
gravity coupled to matter in a chiral gauge \cite{kpz}. They concluded
that the theory is free of anomalies and
solvable for $c_M\le 1$ and $c_M\ge 25$, where $c_M$ is the matter central
charge. Subsequently, their results were rederived in the conformal gauge
by David, Distler and Kawai (DDK)\cite{ddk}.
The structure of the physical Hilbert space was
studied in a series of papers \cite{{hor},{kur},{ito},{liz},{bmp}}.
In all these papers, the physcal Hilbert space was defined as a
cohomology of a BRST charge, which was postulated from the begining,
without a simple and direct relation to a particular action.
The choice of the BRST charge
in \cite{{hor},{kur},{ito}} was motivated by the results of the KPZ analysis,
while the choice of the BRST charge
in \cite{{liz},{bmp}} was motivated by the results of the DDK analysis.
The BRST analysis in the conformal gauge also requires an additional assumption
that the Liouville and the matter sector can be described as free-field
theories with
background charges \cite{bmp}. Another puzzling feature is that the physical
spectrum is like that of a system with finitely many degrees of freedom,
although the starting point is a field theory coupled to gravity.

In this paper we show that all these features of the theory can be easily
understood in the canonical quantization approach, if one starts from the
action for a free scalar field with background charge, coupled to gravity.
In section 2 we describe the canonical structure
and analyze the constraints of such an action.
In section 3 we discuss some general features of the Gupta-Bleuler
and the BRST quantization, which are relevant for our case.
In section 4 we analyze the theory in terms of the $SL(2,{\bf R})$
Kac-Moody variables. A discussion of the issue of
hermiticity is presented, with an emphasis on the matter sector.
In section 5 we introduce variables which transform the theory into a
free-field form, by using the Wakimoto construction for the Kac-Moody
variables.
We then discuss the relation between the
chiral and the conformal gauge spectrum. This is followed by
a discussion about the problem of the complex momentum
of the discrete states and their unitarity.
We present our conclussions in section 6.

\sect{2. Canonical Analysis}

Since we are interested in quantizing 2d gravity coupled
to $c<1$ matter, a natural choice for the classical action is
$$S= -\ha\int_{M} d^2 x \sqrt{-g}( g\sp{\m\n}\pa\sb \m\f\pa\sb \n \f + \a R\f
+ \L ) \quad,\eqno(2.1)$$
where $g\sb{\m\n}$ is a 2d metric, $\f$ is a scalar field, $\a$ is the
background charge, $R$ is the 2d curvature scalar and $\L$ is the cosmological
constant. In the canonical
approach the 2d manifold $M$ must have a toplogy of $\S \times {\bf R}$,
where $\S$ is the spatial manifold and ${\bf R}$ is the real line corresponding
to the time direction. In two dimensions
$\S$ can be either a real line or a circle $S^1$. Since we are interested in
string theory, we will analyze the compact case.
We will label the time coordinate $x^0 = \t$
and the space coordinate $x^1 = \s$.

The configuration space variables are the metric $g\sb{\m\n}(\s,\t)$ and
the scalar field $\f (\s,\t)$. The corresponding canonically conjugate
momenta are defined as
$$ p\sp{\m\n} = {\pa \cl \over\pa \dt{g}\sb{\m\n}} \quad,\quad
   \p = {\pa \cl \over\pa \dt{\f}} \quad,\eqno(2.2)$$
where $\cl$ is the Lagrangian density of (2.1), and $\dt{}$ is the time
derivative.
We define the canonical Poisson brackets as
$$\{g\sb{\m\n}(\s,\t),p\sp{\r\l}(\s^{\prime},\t)\}=\d^{(\m}_{\r}\d^{\n)}_{\l}
\d (\s -\s^{\prime})
\quad,\quad
\{\f(\s,\t),\p(\s^{\prime},\t)\} = \d (\s -\s^{\prime})\, .\eqno(2.3)$$
Since the action (2.1) is invariant under the 2d diffeomorphisms, this
implies that constraints will appear in the canonical formulation.
By performing the canonical analysis \cite{egor},
one can show that (2.2) gives two primary
constraints, corresponding to vanishing of $p\sp{00}$ and $p\sp{01}$.
This means that the corresponding coordinates $g\sb{00}$ and $g\sb{01}$
are non-dynamical. They are the Lagrange multipliers, analogous to the
$A\sb 0$ component of the gauge field in
the case of the Yang-Mills theory.
The secondary constraints are the diffeomorphism constraints
$$\li{G\sb 0 (\s) &= \ha(\f^{\prime})^2 - {2\over{\a^2}}(g_{11}p^{11})^2
-{2\over\a}(g_{11}p^{11})\p
- {{\a}\over2}{g_{11}^{\prime}\over g_{11}}\f^{\prime} + \a\f^{\prime\prime}
- \ha\L g_{11}\cr
      G\sb 1 (\s) &= \p\f^{\prime} - 2g_{11}(p^{11})^{\prime}
- p^{11}g_{11}^{\prime}&(2.4)\cr}$$
which are first class and irreducible, and
obey the diffeomorphism algebra. There are no further constraints, and the
action (2.1) can be rewritten as
$$ S= \int d\s d\t (p^{11}\dt{g}_{11} + \p\dt{\f} - n\sp \m G\sb \m )
 \eqno(2.5)$$
where
$$n^{0}= -{\sqrt{-g}\over g_{11}}\quad,\quad n^1 = {g_{01}\over g_{11}}
\eqno(2.6)$$
are the Lagrange multipliers, imposing the constraints (2.4).

Since we are dealing with a reparametrization invariant system, the
Hamiltonian vanishes on the constraint surface (i.e. it is proportional
to the constraints). Therefore the dynamics is determined by
the constraints only. A straightforward consequence of (2.5) is that (2.1)
does not have any local physical degrees of freedom since
there are as many constraints per space point $\s$, two, as the
number of the configuration variables. This means that there is enough gauge
invariance to gauge away all the $\s$ dependece of $g$ and $\f$, and only
the zero modes may remain. Therefore (2.1) describes a theory without
local physical degrees of freedom. As the subsequent analysis will show,
only certain global degrees of freedom will remain, i.e.
the zero-modes, and in that sense one can think of (2.1) as a topological
theory.

When quantizing a gauge theory, anomalies may appear, which prevent us to
gauge away all of the non-physical degrees of freedom. In our case we have to
examine the quantum theory and see under what conditions the anomalies
cancel, so that the quantum theory remains topological.

\sect{3. Quantum Theory}

In order to quantize
a constrained system, one can adopt the Dirac quantization procedure
\cite{dir}. Given the basic canonical variables $(p_j,q^j)$, we promote
them into hermitian operators $(\hat{p}_j,\hat{q}^j)$,
satisfying the Heisenberg algebra
$$[\hat{p}_j,\hat{q}^k] = -i\d_j^k \quad.\eqno(3.1)$$
A representation of (3.1)
defines the Hilbert space of states $\ch$. The constraint
conditions $G\sb \a (p,q)=0$ are promoted into the operatorial conditions
$$ \hat{G}\sb \a (\hat{p},\hat{q})\sket{\j} = 0 \quad,\eqno(3.2)$$
and the set of states $\sket{\j}$
satisfying (3.2) defines the physical Hilbert space
$\ch^*$. The standard difficulty of the Dirac procedure is
how to define the $\hat{G}_\a$ operators. This difficulty arises because
of the ordering ambiguities. A related problem is that
$\hat{G}_\a$ often
do not form a closed commutator algebra, which is the source of the anomalies.
The anomalies make the conditions (3.2) inconsistent, and one has to use
the Gupta-Bleuler conditions instead, which require that only the expectation
values $\sVEV{\hat{G}_\a}$ vanish. This is usually
equivalent to requiring that only a
subset of $\hat{G}_\a$, which forms a closed subalgebra, anhilates the physical
states. Although a consistent scheme, it is often hard to see what happens
with the anomalies in the Gupta-Bleuler approach. A more suitable approach is
the BRST canonical quantization (for a review and references see \cite{hen}).
In this approach one enlarges the Hilbert space $\ch$ by introducing additional
canonical variables $(c\sp \a , b\sb \a )$, i.e. the ghosts and their canonical
conjugate momenta (antighosts).
Ghosts are of the opposite statistics to $G\sb \a$, and
satisfy
$$ \{ b\sb \a , c\sp \b ] = -i\d_\a^\b \quad,\eqno(3.3)$$
where $\{,]$ is the graded anticommutator. In the space $\ch\otimes\ch_{gh}$,
where $\ch_{gh}$ is a representation of (3.3), one defines an operator
$$ \hat{Q} = c^{\a}\hat{G}\sb \a - \ha i :f\sb{\a\b}\sp \g c\sp \a c\sp \b
b\sb \g: + \cdots \quad,\eqno(3.4)$$
where $f\sb{\a\b}\sp \g$ are the structure constants of the algebra $G$, and
$\cdots$ are determined from the requiriment of nilpotency
$$ \hat{Q}^2 = 0 \quad.\eqno(3.5)$$
Condition (3.5) guarantees the absence of the anomalies in the quantum theory,
and often gives conditions on the free parameters of the theory.
The physical
state conditions (3.2) are replaced with a single condition
$$ \hat{Q}\sket{\F} = 0 \quad.
\eqno(3.6)$$
Only the non-trivial solutions of (3.6) are considered as physical, where
$\sket{\F}= \hat{Q}\sket{\c}$ is trivial. In other words,
the physical Hilbert space is the cohomology of the operator
$\hat{Q}$. For the systems of interest,
the condition (3.6) is
equivalent to the Gupta-Bleuler conditions if
$$\sket{\F} = \sket{\j}\otimes\sket{ghv}\quad,\eqno(3.7)$$
where $\sket{ghv}$ is the ghost-vacuum \cite{gsw}. The states (3.7) form the
``zero'' ghost number cohomology. One can also have physical
states of non-zero ghost number, which correspond to some other consistent
choice of the Gupta-Bleuler conditions. For example,
in the bosonic string case, the usual choice $L_n \j =0,n\ge 0$ corresponds
to $N_{gh} = -\ha$, while $L_n \j = 0,n\ge -1$ corresponds to
$N_{gh}=-\frac32$, where $N_{gh}$ denotes the ghost number and $L_n$ are
the Virasoro generators.
The other possible non-trivial cohomologies arise for $N_{gh} = \ha,\frac32$
\cite{wit}. Note that the consistency of the Gupta-Bleuler conditions
in the case $N_{gh}=-\frac32$ requires vanishing of the string intercept $a$.
Given that $a=(D-2)/24$ \cite{gsw}, this
explains why this cohomology class is empty for the critical string ($D=26$),
while it is non-trivial for a $D=2$ string, which is the case relevant for us.
The BRST formalism is more restrictive than the Gupta-Bleuler formalism,
and the well known example is the bosonic string \cite{gsw},
while a less known example is that of the Siegel superparticle \cite{ilk}.

In our case, the ordering difficulties arise if we use $(g,p)$ and
$(\f,\p)$ as our
basic canonical variables, since the form of $G\sb \m$ is such that
$\hat{G}\sb \m$ will be plagued with ordering ambiguities.
This could be avoided by chosing a more suitable set of canonical variables.
For example, by performing a canonical transformation \cite{{marn},{abd}}
$$\li{\c &= \f - {{\a}\over2} {\rm ln}|g_{11}|\quad,\quad
\p_\c = \p \cr
\x &= {{\a}\over2} {\rm ln}|g\sb{11}| \quad,
\quad \p_\x = {2\over{\a}} g\sb{11}p^{11} + \p \quad,&(3.8)\cr}$$
the constraints become
$$\li{ G\sb 0 &= \ha\p_\c^2 + \ha(\c^{\prime})^2 + \a\c^{\prime\prime}
- \ha\p\sb \x\sp 2 - \ha (\x^{\prime})^2 + \a \x^{\prime\prime}+\ha\L
e^{{2\over\a}\x}\cr
       G\sb 1 &= \p_\c \c^{\prime} + \a\p\sb \c\sp{\prime}
+ \p\sb \x \x^{\prime} - \a\p\sb \x\sp{\prime}\quad.  &(3.9)\cr}$$
If we neglect the cosmological constant term
and the background charges, expressions (3.9) have the same form as the
constraints of a $D=2$ string. In analogy to the string case
we define the left/right movers
$$ h\sb \pm = {1\over \sqrt2}(\p\sb \c \pm \c )\quad,\quad
       j\sb \pm  = {1\over\sqrt2}(\p\sb \x \mp \x ) \eqno(3.10)$$
and redefine the constraints as
$$ T\sb \pm = \ha ( G\sb 0 \pm G\sb 1 )
= \ha h\sb \pm\sp 2 \pm {{\a}\over{\sqrt{2}}}h\sb \pm\sp{\prime}
- \ha j\sb \pm\sp 2 \mp{{\a}\over{\sqrt2}}j\sb \pm\sp{\prime}
+\frac14 \L e^{-{\sqrt2\over\a}(j_+ -j_- )}\quad,\eqno(3.11)$$
which now obey the Virasoro algebra.
In the string case the Virasoro anomaly is $c=D=2$, while $\hat{Q}^2 =0$
requires $c=26$, and therefore the anomaly cannot be removed. In our case
the presence of the background charges and the cosmological term
may change the formula $c=D$ and hence
allow for $c=26$ to be satisfied. Actually, by exploiting the similarity
of $T_{\pm}$ with the energy-momentum tensor of the Liouville theory
(for a review and references see \cite{sib}), one can show that $c=26$ can
be satisfied. However, the $(h,j)$ variables are not convinient for
analyzing the spectrum of the theory. Therefore we are going to look for
a more convinient set of variables.

\sect{4. $SL(2,{\bf R})$ Variables}

The results of the work done in \cite{{isl},{abd}} on the $SL(2,{\bf R})$
symmetry of the induced 2d gravity imply that the
corresponding gauge indipendent variables exist.
Following \cite{abd}, let us introduce non-canonical phase space
variables
$$\li{J\sp + &= {\sqrt{2}\over{g_{11}}} T\sb - +{\L\over 2\sqrt2}\cr
J\sp 0 &= -\left[ g\sb{11}p^{11} + {\a\over2}\left( \p -
{\a\over2}{g_{11}\over
g_{11}^{\prime}}\right)\right] + {1\over\sqrt{2}}x^- J^+ \cr
J\sp - &=- {\a^2\over\sqrt{2}} (g_{11} + 1) + \sqrt{2}x^- J^0 +
{1\over\sqrt{2}}(x^-)^2 J^+\cr
P\sb M &= {1\over\sqrt{2}}\left( \p + \f^{\prime} -
{\a\over 2} {g_{11}\over g_{11}^{\prime}}\right) \quad. &(4.1)\cr}$$
Note that $(J,P\sb M)$ variables cover the whole phase space.
Another important point
is that $x^-$ has to be considered as a constant parameter, indipendent of
$\s$, because otherwise the $J$'s from (4.1) will not be periodic functions
in $\s$, and one could not use the Fouirer modes of $J$ to define the quantum
theory. The $J$'s satisfy an
$SL(2,{\bf R})$ current (Kac-Moody) algebra
$$ \{ J^a (\s\sb 1), J^b (\s\sb 2)\} = f\sp{ab}\sb c J^c (\s\sb 2)
\d (\s\sb 1 -\s\sb 2) +
{\a^2\over2} \h^{ab}\d^{\prime}(\s\sb 1 - \s\sb 2) \quad,\eqno(4.2)$$
where $f\sp{ab}\sb c = 2\e\sp{abd}\h\sb{dc}$ and
$$\h^{ab} =\pmatrix{0 &0 &2\cr 0 &-1 &0 \cr 2 &0 &0\cr}\quad.$$
$P\sb M$ has the Poisson bracket of a free scalar field ($P\sb M =
\pa\sb +\f\sb M $ in the conformal gauge)
$$ \{ P_M (\s\sb 1) , P_M (\s\sb 2 ) \} = \d^{\prime}(\s\sb 1 -\s\sb 2 )
 \eqno(4.3)$$
and $\{J, P\} =0$. Note that the constraints become
$$\li{ J^+ -\l &= 0 \cr T \equiv T\sb g + T\sb M &= 0\quad, &(4.4)\cr}$$
where $\l = {\L\over2\sqrt2}$ and
$$T\sb g = {1\over\a^2}\h_{ab}J^a J^b + (J^0)^{\prime} \quad,
\quad T\sb M = \ha P_M^2 +{\a\over\sqrt2}P_M^{\prime}
 \eqno(4.5)$$
can be interpreted as the gravity and the matter energy-momentum tensors,
respectively.

To construct the Hilbert space $\ch$, we promote $J$ and $P_M$ into hermitian
operators, satisfying the operator version of (4.2-3)
$$\li{ [ J^a(\s\sb 1), J^b(\s\sb 2)] &= if\sb{ab}\sp c J^c
\d (\s\sb 1 -\s\sb 2)
-i{k\over2}\h^{ab} \d^{\prime}(\s\sb 1 - \s\sb 2 ) \cr
 [ P_M (\s\sb 1) , P\sb M (\s\sb 2)] &= i\d^{\prime} (\s\sb 1 -\s\sb 2 )
\quad.&(4.6)\cr}$$
We introduce a new constant $k$, which is different from $\a^2$ due to
ordering ambiguities. It will be determined from the requiriment of anomally
cancelation.
Now one can follow the standard way of constructing $\ch$ as a Fock space
built on the vacuum state anhilliated by the positive Fouirer modes of $J$ and
$P$. Let $f(\s) = \su_n e^{i\e n\s}f_n$, where $\e = \pm 1$,
and let $J_n$, $\a_n^ M$ and $L_n$ denote the Fouirer
modes of $J$, $P_M$ and $T$, respectively. We represent the Fock space vacuum
as $\sket{j,m}\otimes\sket{p_M}$, where $\sket{j,m}$ is the vacuum for the
Kac-Moody sector, while $\sket{p_M}$ is the vacuum for the matter sector.
The Kac-Moody vacuum states satisfy
$$\li{J^a_n \sket{j,m} &= 0 \quad,\quad n\ge 1 \cr
  J^a_0 \sket{j,m} &= j^a \sket{j,m}\quad,&(4.7)\cr}$$
where the last condition means that the vacuum states form an $SL(2,{\bf R})$
representation. Unitary $SL(2,{\bf R})$ representations are infinite
dimensional
(since $SL(2,{\bf R})$ is a non-compact group), and can be labeled with a
complex number $j$, which can take the following values
$$ j=\ha + ir \quad,\quad r\in {\bf R} \quad{\rm or}\quad
0<j<1 \quad{\rm or}\quad j \quad{\rm is\, a\, half-integer}\quad,\eqno(4.8) $$
where $j(j-1)$ is an eigenvalue of $j^a j_a$, while the second label
$m\in {\bf Z}$, is an eigenvalue of $j^0$ \cite{vil}.

The matter vacuum satisfies a $U(1)$
version of (4.7)
$$\li{\a_n^M \sket{p_M} &= 0 \quad,\quad n\ge 1 \cr
  \a_0^M \sket{p_M} &= p_M \sket{p_M}\quad.&(4.9)\cr}$$
$\sket{p_M}$ is the usual momentum state, so that $p_M$ is real and
continious eigenvalue. The $\a_n^M$ modes satisfy
$$[\a_n^M ,\a_m^M ] = -\e n\d_{n+m,0}\quad,\eqno(4.10)$$
which gives for the matter central charge
$$ c_M = 1 -\e 12 Q_M^2 \quad,\eqno(4.11)$$
where $Q_M = \sqrt{\p}\a$.
Hence we will chose $\e = 1$ in order to have $c_M < 1$.
If we want the $L_n$'s to satisfy the usual Virasoro algebra
$$ [L_n ,L_m ] = (n-m)L_{n+m} + A_n \d_{n+m,0}\quad,\eqno(4.12) $$
where $A_n$ is the anomally,
we have to change the sign of the $L_n$'s coming from (4.4), i.e. define
$-T$ as our energy momentum tensor. Hence
$$ -T = {1\over 2\p}\su_n L_n e^{in\s} \quad,\eqno(4.13)$$
where
$$L\sb n\sp g ={1\over k+2}\su_m \h_{ab} J_{n-m}^a J_m^b -in J_n^0 \eqno(4.14)
$$
and
$$ L\sb n\sp M = -\ha \su_m \a^M_{n-m}\a^M_m -inQ_M \a^M_n \quad.\eqno(4.15)$$
Note that the factor ${1\over\a^2}$ in
the classical expression for $T\sb g$ in (4.5) has become $-{1\over k+2}$.

The BRST quantization requires the enlarged Hilbert space
$\ch\otimes\ch_{gh}$, and the whole set up is
equivalent to the starting point of the
chiral gauge analysis of \cite{{hor},{kur}}. The only difference is that our
expressions are gauge indipendent. Our choice of the Kac-Mody modes
is the same as that of \cite{hor}, while the choice of the matter modes
is different from \cite{{hor},{kur}}. Their modes, which we denote as
$\bar{\a}_n^M$, are related to ours as
$$ \a_n^M = \bar{\a}_n^M + i Q_M \d_{n,0}\quad,\eqno(4.16)$$
and their relation to our modes is analogous to the relation of the Kac-Moody
modes of \cite{kur} to the Kac-Moody modes of \cite{hor}.
The barred modes are often used in conformal field theory, and
arise from mapping the cylinder $S^1\times{\bf R}$ onto the complex plane.
The corresponding mode expansion is given by
$$\bar{P}_M (z) = {1\over\sqrt{2\p}}\su_n \bar{\a}_n^M z^{-n-1}\quad,
\eqno(4.17)$$
where $z\bar{P}_M (z)$ coincides with $P_M (\s)$ for $z=e^{-i\s}$.

Note that our choice of representation for the matter $L_n$'s is not the one
used in the conformal field theory (CFT). Namely, in order to have the usual
hermitian conjugacy rules
$$\a_n\sp{\dag}= \a_{-n} \,\to \, L_n\sp{\dag} = L_{-n} \eqno(4.18) $$
and $c_M < 1$, we had to introduce
``negative" $L_n$'s, given by (4.15). As a consequence, the matter Fock
space has a negative norm, since $\e =1$ in (4.10).
This is not a problem, since the positivity of the
norm is only required for the physical Fock space. One could have chosen the
CFT representation \cite{fel}, where $L_n$'s are ``positive"
$$ L\sb n\sp M = \ha \su_m \a^M_{n-m}\a^M_m + n Q_M \a^M_n \eqno(4.19)$$
and the $\a$'s satisfy (4.10) with $\e=-1$.
But then in order to have $c_M <1$, the background charge has to be imaginary,
and the $L_n$'s will not be hermitian under the usual scalar product
represented by the rules (4.18).
A modified scalar product can be introduced, which acts on $F^* \times F$,
where $F^*$ is the dual of the matter Fock space $F$ \cite{fel}.
The duality relation has a property that
$F_{2Q-p}^*$ is isomorphic to $F_p$, where $p$ is the momentum of the vacuum
state. However, we are going to
use the representation (4.15), to which we are going to refer as the string
representation,
since it is more convinient for our approach.

The BRST charge can be constructed from the equation (3.4)
$$ \hat{Q} = c\sb 0 ( L\sb 0 - a ) + \su_{n \ne 0} c\sb n  L\sb{-n} +
\su_{n} c^+_{n}J^+_{-n} + \cdots \quad,\eqno(4.20)$$
where $a$ is the intercept. The
nilpotency condition requires vanishing of the total central charge,
which includes the matter, Kac-Moody and the ghosts contributions
$$ c_M + {3k\over k + 2 } - 6k - 26 - 2 = 0 \quad,\eqno(4.21)$$
and the intercept must satisfy
$$ a = 1 + {k\over4} + \ha Q_M^2 \quad.\eqno(4.22)$$
Expression (4.22) differs from the corresponding expression in \cite{hor}
because we used the string representation
modes $\a_n$. Note that the equation (4.21) implies
$$ k+3 = {1\over12}\left(c_M - 1 \pm \sqrt{(1-c_M)(25-c_M)} \right)
\quad,\eqno(4.23)$$
and therefore $k$ is real if $c_M\le 1$ or $c_M\ge 25$, which justifies our
choice $\e =1$.

Only the zero ghost number cohomology is non-trivial \cite{{hor},{kur}},
which corresponds to the usual Gupta-Bleuler conditions
$$ ( L_n - a \d_{n,0})\sket{\j} = 0 \quad,\quad (J_n^+ -\l\d_{n,0})\sket{\j}
= 0 \quad,\quad n\ge 0 \quad.\eqno(4.24)$$
It consists of the vacuum states of the Fock space $\ch$
$$ \sket{\j\sb 0} = \sket{j}\otimes\sket{p\sb M} \quad,\eqno(4.25)$$
where $\sket{j}$ satisfies $j^+\sket{j}= \l \sket{j}$.
$j$ and $p_M$ are related through the ground state
on-shell condition
$$(L_0 - a )\sket{\j\sb 0} = 0 \to 1 = {j(j - 1 )\over k + 2} - {k\over 4}
- \D(p_M)  \quad.\eqno(4.26)$$
$\D(p_M)$ plays the role of the matter conformal dimension, and can be
expressed as
$$ \D(p_M) = \ha ( p^2_M + Q^2_M ) = \ha \bar{p}_M (\bar{p}_M + 2i Q_M )
\quad,\eqno(4.27)$$
where $\bar{p}$ is the eigenvalue of the CFT mode $\bar{\a}_0^M$.
This coincides with the usual formula for $\D(p_M)$ if
$\bar{p}$ is imaginary. Neglecting for the moment
the problem of the imaginary momenta,
which we are going to discuss in the next section, the
formula (4.9) coincides with the expression given in \cite{hor}.
Therefore the quantum analysis confirms our classical picture of only the
zero modes being physical.

\sect{5. Free-field Variables}

The exsistence of the $SL(2,{\bf R})$ variables (4.1) is a strong indication
that the conformal gauge
variables used in \cite{{liz},{bmp}} should also have a gauge indipendent
realization. The results of Itoh's work in the chiral gauge \cite{ito}
imply that the new variables can be determined from the
Wakimoto's construction \cite{wak}. Let us
introduce three new variables $\b$, $\g$ and $P\sb L $ such that
$$ \li{J^+ (\s) &= \b (\s)\cr
J^0 (\s) &=\b(\s)\g(\s) + k_1 P\sb L (\s)\cr
J^- (\s) &= \b(\s)\g^2(\s) + 2k_1 \g(\s) P\sb L (\s) +
k_2 \g^{\prime} (\s) \quad,&(5.1)\cr}$$
and require
$$ \{\b (\s\sb 1 ) ,\g (\s\sb 2)\} =  \d (\s\sb 1 -\s\sb 2) \quad,\quad
\{P\sb L (\s\sb 1 ) , P\sb L (\s\sb 2)\} = -\d^{\prime} (\s\sb 1 -\s\sb 2)
\quad,\eqno(5.2)$$
whith the other Poisson brackets bieng zero.
{}From the requiriment that the expressions (5.1) satisfy the Poisson algebra
(4.2), one can easily see that
$$ k_1 = {\a\over\sqrt2}\quad,\quad k_2 =\a^2 \quad.\eqno(5.3)$$
In terms of the new variables $T_g$ becomes
$$ -T_g = -\b^{\prime}\g + \ha P^2_L + {Q_L\over\sqrt{2\p}}P^{\prime}_L \quad,
\eqno(5.4)$$
where $Q_L = -\sqrt{\p}\a$. In the quantum case, $\b$, $\g$ and $P_L$
become hermitian operators, satisfying
$$ [\b (\s\sb 1 ) ,\g (\s\sb 2)] = i \d (\s\sb 1 -\s\sb 2) \quad,\quad
[P\sb L (\s\sb 1 ) , P\sb L (\s\sb 2)] = -i \d^{\prime} (\s\sb 1 -\s\sb 2)
\quad.\eqno(5.5)$$
The quantum analog of (5.1) is
$$ \li{J^+ (\s) &= \b (\s)\cr
J^0 (\s) &=:\b(\s)\g(\s): + k_1 P\sb L (\s)\cr
J^- (\s) &= :\b(\s)\g^2(\s): + 2k_1 \g(\s) P\sb L (\s) +
k_2 \g^{\prime} (\s) \quad,&(5.6)\cr}$$
where now $k_1$ and $k_2$ have acquired new quantum values
$$k_1^2 ={1\over\a_+^2}=-(k+2) \quad,\quad k_2 =-k \quad, \eqno(5.7)$$
due to the normal ordering effects.
The normal ordering in (5.6) is with respect to the vacuum $\sket{vac}$
$$ \b_n \sket{vac} = \g_n \sket{vac} = 0 \quad,\quad n\ge 1 \quad,\eqno(5.8)$$
where $\b_n$ and $\g_n$ are the Fouirer modes of $\b$ and $\g$. The expression
for $T_g$ retains the classical form (5.4), with the appropriate normal
ordering. However, $Q_L$ acquires the quantum value
$Q_L = - {2\over\a_+}-\a_+$.

The constraints can be now written as
$$ J^+ =\b = 0 \quad,\quad -T = -\b^{\prime}\g + \ha P^2_L +
{Q_L\over\sqrt{2\p}} P^{\prime}_L
 - \ha P^2_M - {Q_M\over\sqrt{2\p}} P^{\prime}_M  = 0 \quad,\eqno(5.9)$$
where $\b$ in (5.9) is shifted by the constant $\l$.
Vanishing of $\b$ means that we can drop that variable, together with its
canonically conjugate variable $\g$,
and we are left with $P\sb L$ and $P\sb M$ variables,
obeying only one constraint
$$ -T \approx T\sb{cf} \equiv \ha P_L^2 + {Q\sb L\over\sqrt{2\p}} P_L^{\prime}
 - \ha P^2_M - {Q_M\over\sqrt{2\p}} P^{\prime}_M =0 \quad.\eqno(5.10)$$
But this is precisely the starting point of the conformal gauge analysis
\cite{{liz},{bmp}}. The only difference is that the expression (5.10) is
gauge independent, and in the conformal gauge reduces to the expression
used in \cite{{liz},{bmp}}.

If we introduce a notation
$$ X^{\m}= (X^L , X^M )\quad,\quad X\cdot Y = \h_{\m\n}X^\m Y^\n \quad,
\quad \h_{\m\n} = \pmatrix{1 &0\cr 0 &-1\cr} \quad,\eqno(5.11)$$
then
$$ L_n = \ha\su_m :\a_{n-m}\cdot\a_{m}: + inQ\cdot\a_n \quad.\eqno(5.12)$$
The BRST charge is then given by the usual expression
$$ \hat{Q} = \su_n c_{n} L_{-n} + \ha \su_{m,n} (m-n):c_m c_n b_{-m-n}:
-c_0 a \quad.\eqno(5.13)$$
The normal ordering is with respect to the vacuum $\sket{vac}=\sket{p}\otimes
\sket{0}$
$$ \a_n\sket{vac}= c_n \sket{vac} = b_{n}\sket{vac} = 0 \quad,\quad n \ge 1
\quad,\eqno(5.14)$$
where $\sket{p}$ is the $\a$-modes vacuum ($\a_0\sket{p}=
p\sket{p}$), while $\sket{0}$ is the ghost vacuum, satisfying $b_0\sket{0}=0$
(the other possibility $c_0\sket{0}=0$ gives symmetric results).
Nilpotency of $\hat{Q}$ implies
$$ Q^2 = Q^2_L - Q^2_M = 2 \quad,\quad a =0 \quad.\eqno(5.15)$$

The results of the BRST analysis in \cite{{liz},{bmp}} can be now understood
in the following way. The zero-ghost number cohomology corresponds to the
usual Gupta-Bleuler conditions
$$ L_n \sket{\j} = 0 \quad,\quad n\ge 0 \quad,\eqno(5.16)$$
where $\sket{\j}$ belongs to the $\a$-modes Fock space $F(\a)$.
Clearly, the ground state $\sket{p}$ is a solution of (5.16) if
$$p^2 =p^2_L - p^2_M = 0\quad.\eqno(5.17)$$
In terms of the CFT modes, (5.17) translates into
$$ \D(\bar{p}_L) - \D(\bar{p}_M) = 1 \quad.\eqno(5.18)$$
Note that (5.17)
is equivalent to the $SL(2,{\bf R})$ conditon (4.26), and $\sket{p}$ is
the same as $\sket{j,p_M}$. This implies the relation
$${j(j-1)\over k+1} - {k\over4}=\ha Q_M^2 + \ha p_L^2 \ge 0 \quad,\eqno(5.19)$$
which is satisfied if $j$ belongs to any of the continious series of
representations from (4.8), and if $k$ is given by the negative root of (4.23).
The negative root is taken because then $k+3$ coincides with the string
susceptibility coeficient $\g_{str}$ \cite{ddk}.
If $j$ belongs to the discrete series, then (5.19) is satisfied for
$j_-(Q_M) \le j \le j_+ (Q_M)$, where $j_{\pm}(Q_M)$ are the roots of
the equation (5.19).

As far as the excited states are concerned, the results of the BRST analysis
imply that they are physical only for discrete values of the momenta
\cite{{liz},{bmp}}. Furthermore, when translated into our conventions, these
discrete values of the momenta are purely imaginary
$$\li{p_L &= \frac{i}2 (r +s) Q_L - \frac{i}2 (r-s) Q_M \cr
p_M &= \frac{i}2 (r -s) Q_L - \frac{i}2 (r+s) Q_M \quad,&(5.20)\cr}$$
where $r,s \in {\bf Z}$, and $rs$ is the excitation level number.

The states in the $\pm1$ cohomology sector have only discrete
values of the momenta. They are of the form
\cite{bmp}
$$\sket{\j}\otimes b_{-n}\sket{0} \quad {\rm or} \quad
\sket{\j}\otimes c_{-n}\sket{0} \quad,\quad n\ge 1\quad,\eqno(5.21)$$
where $\sket{\j} \in F(\a)$. Absence of the continious momentum states in this
case can be understood on the example of $\sket{\j}\otimes b_{-1} \sket{0}$,
since then (3.6) implies $L_n \sket{\j}=0$ for $n\ge -1$, which
for the ground state $\sket{p}$ implies $p=0$. Similarly to the zero ghost
number case, the excited states are physical only for
complex discrete values of the momenta, given by the equation (5.20).

The fact that all discrete
states have complex momenta explains why they were not
found in the analysis of \cite{{hor},{kur},{ito}}, since they are not defined
in the standard framework.
According to the standard construction
of the free-field Fock space $\ch$, the discrete states do not even belong to
$\ch$, because of the complex momentum. However, there are strong indications
that the discrete states are physical \cite{poly}, and
in order to incorporate them into a Hilbert
space, one has to find another free-field realization of the Fock space $\ch$.
One can see the difficulty in doing this by
considering the zero-modes sector, where a representation of the Heisenberg
algebra has to be constructed. The usual momentum states $\sket{p}$ are
constructed as
$$ \sket{p} = e^{ip\hat{q}}\sket{p=0} \quad.\eqno(5.22)$$
The states (5.22) are $\d$-function normalizable for Im$p=0$, while otherwise
cannot be normalized. According to the Stone-von Neumann theorem \cite{sto},
Im$p=0$ is the only inequivalent unitary
irreducible representation of the Heisenberg algebra, which implies that
complex momentum states are not unitary.
A possible resolution of this problem may be in the fact that
the Stone-von Neumann theorem applies to the case when
$-\infty <q<+\infty$. When $0\le q<+\infty$, a case relevant for the
components of the metric, then $\hat{q}$ is not hermitian with respect to
the usual scalar product, and
the Stone-von Neumann theorem does not apply any more.
This will require a further investigation, in particular a careful treatment
of the range of $q_L$, a coordinate canonically conjugate to the zero-mode
of $P_L$.

Note that Sieberg has found the same phenomenon, i.e. non-normalizability of
the complex momentum states, in the Liouville theory approach to 2d quantum
gravity \cite{sib}. This indicates a strong connection between our model and
the Liouville theory. In order to better understand this connection, a study
of our theory in terms of the variables defined by the equation
(3.8) will be suitable.

\sect{6. Conclusions}

We have demonstrated that the conformal gauge results of DDK
can be derived in the
gauge indipendent way. To do this, we have used the Dirac quantization
procedure, which is gauge indipendent and therefore convinient for such a
task. In order to obtain the freee-field variables $(\b,\g,P_L,P_M)$,
we went through a series of transformations
$$(g,p,\f,\p) \to (J^a , P_M) \to (\b,\g,P_L,P_M )\quad.\eqno(6.1)$$
Note the importance of
the sequence (6.1), since it implicitly defines the $ (g,p,\f,\p) \to
(\b,\g,P_L,P_M)$ transformation. The $J$ variables are also
important for understanding the $c_M <1$ constraint, as well as
for the simple calculation of the $Q_L$ renormalization.

Given the free-field variables, one can use the results of the BRST
analysis \cite{{liz},{bmp}} to obtain the physical spectrum of the theory.
The analysis of the spectrum confirms the classical picture of only
the zero-modes of the gravity and the matter sector propagating, which can be
described as states of a $D=2$ massless relativistic particle. Existence of
the discrete states means that the massive states are not completely pure gauge
states, and can be physical for specific discrete values of the momenta.
However, incorporating
the discrete states into a Hilbert space is still an open question,
due to their complex momentum,
and further work along the lines suggested in section 5 is necessary.

Our results imply the following physical picture: 2d guantum gravity coupled
to a scalar field is described by a
Liouville-like theory if one uses the variables defined by the equation
(3.8). The quantum theory can be transformed
into a free-field form for $c_M \le 1$ or
$c_M\ge 25$. For these values of $c_M$ the quantum theory retains its
classical topological features. Note that in the case when the scalar field
describes a minimal CFT, then the theory looks even more topological, since
then $\D(\bar{p}_M)$ can take only discrete
values, and one is left with only discrete momentum states. This implies that
the effective field theory describing the interactions among these states is
zero-dimensional, which explains why the zero-dimensional
matrix models can be used to describe the minimal models coupled to gravity.
Formulating the interacting theory in the canonical approach
can be done in a string field theory framework.

The $c_M =1$ case does not follow from the canonical analysis of (2.1) with
$\a =0$. The $\a =0$ case is just a $D=1$ bosonic string theory.
According to the no-ghost theorem \cite{thorn}, if
$D<26$ then there are $D-1$ physical degrees of freedom per space point $\s$.
For the $D=1$ case this means that only the zero modes are propagating,
which formaly agrees with the $c_M =1$ path-integral result.
However, $\hat{Q}^2 \ne 0$ in the canonical treatment of the
$D=1$ string, and conformal anomaly is
present. A way to reconcile these results
is that $c_M =1$ case with $\hat{Q}^2 =0$ can be obtained in the canonical
approach from a $D=2$ string with a dilaton coupling.

As far as the supersymmetric case is concerned,
we expect that the canonical treatment of the supersymmetric generalization
of the action (2.1) will give the results analogous to the bosonic case,
i.e. that only the zero modes of the super-matter and the super-Liouville
sector will propagate. This would rigorously prove the results of the
super-conformal gauge BRST analysis \cite{susy}.

\sect{Acknowledgements}

I would like to thank S. Thomas,
W. Sabra, K. Stelle, C. Hull and G. Papadopoulos for helpful discussions.

\end{document}